\documentclass[12pt]{article}

\usepackage{amsfonts}
\usepackage{latexsym}
\usepackage{amsmath,amssymb}
\usepackage{verbatim}
\numberwithin{equation}{section}
\def\half{{\textstyle{1\over2}}}
\def\p{\partial}

\newcommand{\bea}{\begin{eqnarray}}
\newcommand{\eea}{\end{eqnarray}}
\newcommand{\be}{\begin{equation}}
\newcommand{\ee}{\end{equation}}
\newcommand{\ba}{\begin{align}}
\newcommand{\ea}{\end{align}}

\def\half{{1\over2}}
\def\ft#1#2{{\textstyle{{\scriptstyle #1}\over {\scriptstyle #2}}}}

\def\nn{\nonumber}
\def\sst#1{{\scriptscriptstyle #1}}

\def\rr{{\sst {\rm RR}}}
\def\cF{{\cal F}}
\def\0{{\sst{(0)}}}
\def\1{{\sst{(1)}}}
\def\2{{\sst{(2)}}}
\def\3{{\sst{(3)}}}
\def\4{{\sst{(4)}}}
\def\5{{\sst{(5)}}}
\def\6{{\sst{(6)}}}
\def\7{{\sst{(7)}}}
\def\8{{\sst{(8)}}}

\def\p{\partial}

\def\cal#1{\mathcal{#1}}

\def\ds4{-\omega_+\omega_-+\sigma_1^2+\sigma_2^2}

\def\S2{\Sigma_2}
\def\O2{\Omega_2}

\def\domega2{\sigma_1^2+\sigma_2^2}

\def\half {{1\over2}}

\def\Sg2{{1\over2}\omega_+\wedge\omega_-}

\def\domega2{\sigma_1^2+\sigma_2^2}

\def\ds4{-\omega_+\omega_-+\sigma_1^2+\sigma_2^2}

\def\Q{{\cal Q}}

\begin{document}
\begin{titlepage}
\today
\vskip 5 cm
\begin{center}
{\large{\bf{D-brane Construction of the 5D NHEK Dual
}}}\vskip 0.5in
{Wei Song
and Andrew Strominger}
\vskip 0.3in
{\it Center for the Fundamental Laws of Nature,
Harvard University\\
Cambridge, MA, 02138

}

\end{center}

\vskip 0.6cm

\begin{abstract}
Extremal but non-supersymmetric charged black holes with $SU(2)_L$ spin in IIB string theory compactified to five dimensions on $K^3\times S^1$ are considered. These have a  near-horizon or NHEK region with an enhanced $SL(2,R)_L$ conformal symmetry. It is shown that the NHEK geometry has a second, inequivalent, asymptotically flat extension in which the radius of the $S^1$ becomes infinite but the radius of the angular circles of $SU(2)_L$ orbits
approach a constant.  The asymptotic charges associated to the second solution identify it as a  5D D1-D5-Taub-NUT black string with certain nonzero worldvolume charge densities, temperatures and chemical potentials.  The dual of the NHEK geometry is then identified as an IR limit of this wrapped brane configuration.
\end{abstract}

\vspace{3.0cm}

\end{titlepage}

\setcounter{tocdepth}{2}
\tableofcontents
\section{Introduction}It has been conjectured \cite{ghss} that extreme astrophysical Kerr black holes with spin $J$  are dual to 2D conformal field theories with central charge $c=12J$. In the real world, we cannot expect to know the exact form of the CFT, as that would be equivalent to knowing all the laws of physics to the Planck scale and beyond. Rather, a compelling match has been found between universal properties of the CFTs and universal properties of Kerr black holes. For a recent review see \cite{cargese}.

  On the other hand it is nevertheless useful, in order to better understand the nature of the proposed real-world Kerr/CFT correspondence, to have a toy model in which the exact form of the CFT might be found.  Such a model was recently constructed \cite{Guica:2010ej} by embedding certain 5D charged spinning black holes in string theory.  At the maximal allowed value of the charge, the near horizon NHEK geometry is AdS$_3 \times S^2$, and standard string theoretic methods
were used to identify the dual in terms of (a long string of) the wrapped D1-D5-Taub-NUT CFT.  This result, as well as the properties of linear perturbations around maximality, all agree with expectations from the Kerr/CFT conjecture.

The issue of finite deviations from maximality, where the near-horizon geometry does not contain an AdS$_3$ factor, was not substantially addressed in \cite{Guica:2010ej}. In the present paper, we go beyond perturbative excursions from maximality and identify the
dual for finite deformations as the IR limit of a wrapped D1-D5-Taub-NUT string with certain non-zero charge densities and temperatures. This is acomplished by constructing a 5D black string solution whose near-horizon geometry is locally identical to the 5D NHEK. The full black string geometry  differs from the full black hole geometry in part because the angular circle does not grow in size at large radius.  The microscopic D-brane configuration corresponding to the black string geometry is  straightforward to read off from the asymptotic values of the electric and magnetic fields and is presented in table 1 below.  The dual of the NHEK geometry, for generic charges and spins is thereby identified as an IR limit of this D-brane configuration.\footnote{ The  precise nature of the flow to the IR (i.e. how charge densities are scaled etc.) can be determined from the radial dependence of the geometry. The nature of the IR limit here is clearly somewhat exotic because of the warping of the near-horizon AdS$_3$, potentially related to a so-called dipole deformation of the IR CFT \cite{bergman}. We do not address any of these interesting issues herein but hope to return to them elsewhere.}

This paper is organized as follows.  In section 2 we review the 5D charged black hole solutions described in \cite{Guica:2010ej}, and introduce a slight generalization in which one of the charges and the asymptotic scalar are varied.  In section 3 we construct new, asymptotically flat black string solutions of the same theory.  The most general rotationally-invariant black string solution of 5D minimal supergravity does not have a warped near-horizon region \cite{Compere:2010fm}: a second $U(1)$ charge is needed for this purpose. 
Using the asymptotic values of the various fields at spatial infinity we  compute and exhibit (see table 1) all the various magnetic charges, electric charge densities and ADM energy-momentum of the black string. This information enables us to identify the corresponding wrapped D-brane configuration. In section 4 we show that the near-horizon region of the 5D black holes of section 2 are a finite-temperature identification of the near-horizon region of the black strings of section 3, thereby identifying the specific collection of wrapped D-branes  whose IR limit is dual to the 5D charged NHEK geometry.

\section{5D extremal spinning black holes}

In this section we describe  non-supersymmetric, extremal, charged spinning black holes which arise in the compactification of IIB string theory to 5 dimensions on $K3\times S^1$ and their NHEK limits.
\subsection{The full solution}
We consider the 6-dimensional Einstein-frame effective action
\be \label{act}
\mathcal{S}_{6B}={1 \over 8 \pi ^3}\int d^6x\sqrt{-g}\bigl(R-{1\over12}(F^{RR}_{(3)})^2 \bigr).
\ee
This truncation of the low energy action without a dilaton is consistent when we make the additional self-duality restriction
\be F^{RR}_{(3)}=*F^{RR}_{(3)}.\ee
We  further specialize to  black hole solutions of this action in a Kaluza-Klein (KK) compactification to $D=5$ which carry only $SU(2)_L$ angular momentum $J_L$. These restrictions are made in order to illustrate the basic concepts in the simplest possible setting.
More general solutions can be obtained by U-duality.

The KK compactification of (\ref{act}) to 5 dimensions has one $U(1)$ gauge field descended from $ F^{RR}_{(3)}$, a second KK $U(1)$ gauge field and a scalar parametrizing the radius of the $S^1$.  A careful presentation of the relation between 5D and 6D variables in this compactification can be found in \cite{Cremmer:1997ct, Duff:1998cr}, but we mostly employ the simpler 6D variables throughout. The $J_R=0$ extreme black hole solutions are labeled by three internal parameters: the mass (or $J_L$) and two charges.  The asymptotic constant value of the scalar is a fourth external parameter. More general 5D black holes with more nonzero charges can be found in \cite{Cvetic:1996xz}, whose six dimensional embedding can be found in \cite{Giusto:2004id}.

We first review the subset of  solutions studied in \cite{Guica:2010ej} characterized by the two parameters $a$ and $\delta$, and then present the simple two-parameter generalization. $a$ and $\delta$ are related to the mass $M$, graviphoton charge $Q$ and spin $J_L$ by
\bea
M &=& {6a^2} \cosh 2\delta
\\
J_L  &=&4 a^3\,(c^3 + s^3 ),
\\
Q  &=& 4a^2sc
\eea
where $c = \cosh \delta$ and $s = \sinh \delta$. These relations imply
\be
 Q^3 \leq J_L^2 \;,\;\;\;\;\;   \left(\frac{M}{3} - \frac{Q}{2} \right)^2
\left(\frac{2M}{3}+ 2Q\right) =J_L^2. \label{ppo}
\ee
The metric is
\bea
\label{nhekstring}
ds_6^2  &=& -{\left(\hat r^2+a^2(1-4 c^2) \right)\over \Sigma}d\hat{t}^2+{d\hat u^2}  +{4a^2\sinh 2\delta\over \Sigma}\,
d\hat{t} d\hat u +
 \Sigma\left( {\hat{r}^2 d\hat{r}^2\over ( \hat{r}^2-a^2)^2} + \frac{d\theta^2}{4}\right)\notag\\
& & + \frac{\Sigma}{4}(\,d\hat{\psi}^2 +  d\hat{\phi}^2 + 2 \cos\theta \,d\hat{\phi} \,d \hat \psi)
 + {a^4\over  \Sigma}(d\hat{\psi} + \cos\theta \,d\hat{\phi})^2 \notag \\
& & - {4a^3\over \Sigma}\left( (c^3+ s^3)\,d\hat{t} + (s^2 c + c^2 s)\,d\hat u\right)( d\hat{\psi} + \cos \theta \,d\hat{\phi})
\eea
where $\Sigma \equiv \hat r^2+a^2(1+4 s^2) $ and \be\label{id} \hat u \sim \hat u +2\pi m
\ee
is the coordinate of the unit-radius KK $S^1$. Defining the one-form
\be
 A= \frac{2a^2  \sinh 2\delta }{ \Sigma} \left( d\hat{t} - \half a e^\delta (d\hat{\psi} + \cos\theta \, d\hat{\phi}) \right)
\ee
the RR three-form is given by the manifestly self-dual expression
\be
\label{6ft}
 F^{RR}_{(3)}   =     -dA\wedge(d\hat u+A)-*\bigl(dA\wedge(d\hat u+A)\bigr).
         \ee

         This two-parameter family of solutions is easily embedded in the larger four- parameter family, in which the two $U(1)$ charges and scalar field are separately varied. We simply deform the identification (\ref{id}) by the parameters $\lambda$ and $R$ to
\be \label{zidd} \hat u \sim \hat u +2\pi mR \cosh \lambda,~~~\hat t\sim \hat t -2\pi m R\sinh \lambda,
\ee
which does not introduce any singularities.  To understand the asymptotic geometry we
boost to the primed  coordinates
\be \hat u' =  \cosh \lambda \hat u +\sinh \lambda \hat t ,~~~\hat t '=\cosh \lambda \hat t +\sinh \lambda \hat u,\ee
which have the canonical identification
\be \label{isdd} \hat u' \sim \hat u' +2\pi mR,~~~\hat t'\sim \hat t' .
\ee
Hence the asymptotic geometry differs only by the radius $R$ of the KK $S^1$, previously set to unity.

\subsection{The NHEK limit}

In this subsection we describe the near-horizon limit, following \cite{Dias:2007nj,Bredberg:2009pv,Guica:2010ej}.
Define
\begin{equation}
t = {\Omega_L\over 2a^2} \hat{t} \epsilon \ , \quad \quad   r = {\hat{r}^2 - a^2\over \epsilon} \ ,
\quad \quad y = {2 \pi T_Q}(\hat u + \Phi \hat{t}) \ ,
\ee
\be
\psi = \hat{\psi} - \Omega_L \hat{t}-{2J_L\over Q^2}(\hat u + \Phi \hat{t}) \ , \quad \quad
\phi = \hat{\phi}  \label{nhlim6D }
\end{equation}
where
\bea
{T_Q}&\equiv& {c^3-s^3 \over
4\pi a s^2c^2}={\sqrt{J_L^2-{Q^3 }} \over  \pi Q^2},\\
\Omega_L&\equiv&{1 \over a(c^3-s^3)},\\
\Phi&\equiv&{c^2s-s^2c \over c^3-s^3}.
\eea
The near-horizon metric is then the $\epsilon \to 0$ limit of (\ref{nhekstring})
\bea\label{ff}
{12\over M}ds^2  &=& -r^2 dt^2 + {dr^2\over r^2} + \gamma(dy+rdt)^2 + \gamma(d\psi + \cos\theta d\phi)^2 \\
& & + 2 \alpha \gamma (dy + r dt)(d \psi + \cos\theta d\phi) + d\theta^2 + \sin^2\theta d\phi^2\notag
\eea
where the deformation parameters
\bea\label{squashparam}
\alpha = {2 \cosh 2 \delta \over 1+ \cosh^2 2 \delta},\quad
\gamma = 1 + {1\over \cosh^2 2 \delta} \ .\label{alphagamma}
\eea
are related by $M\alpha\gamma=12a^2$.

In terms of the $SL(2,R)_L \times SU(2)_R$ invariant forms,
\bea
\sigma_1 &=& \cos\psi d\theta + \sin\theta \sin\psi d\phi\\
\sigma_2 &=& -\sin\psi d\theta + \sin\theta \cos\psi d\phi\notag\\
\sigma_3 &=& d\psi + \cos\theta d\phi\notag\\
w_{\pm} &=& -e^{\mp y}rdt \mp e^{\mp y}dr/r\notag\\
w_3 &=& dy + r dt \ ,\notag
\eea
the metric can be written
\bea\label{nicemetric}
{12\over M}ds^2 &=& -w_+ w_- + \gamma w_3^2+ \sigma_1^2 + \sigma_2^2  + \gamma \sigma_3^2 +2 \alpha \gamma w_3 \sigma_3\ .
\eea
 The gauge field $A$ reduces to
\be
d\hat u+A= -{a \over 2}\tanh 2 \delta \bigl(e^\delta\sigma_3+e^{-\delta}w_3\bigr).
\ee
$ F^{RR}_{(3)}$ follows from $A$ via (\ref{6ft}) as
\be \label{hf}
 F^{RR}_{(3)} = { \frac{Q}{4} } \left( \sigma_1 \wedge \sigma_2 \wedge \sigma_3 + \frac{1}{2} \, w_+
\wedge w_-\wedge w_3  + {\rm sech} 2 \delta \, ( \sigma_1 \wedge \sigma_2 \wedge w_3 +
\frac{1}{2} \, w_+ \wedge w_-\wedge
\sigma_3 )\right)
\ee
The $(y,\psi )$
identifications (\ref{zidd}) are
    \be \label{id} y \sim y+ 4\pi^2 T_R m,~~~~~\psi \sim \psi+2\pi \Theta m+4\pi n, \ee for any integers $(m,n)$.  Here we have defined
\bea T_R&=&{RT_Q }({ }\cosh \lambda- \Phi\sinh \lambda),\\
\Theta&=&- {2 J_LR \over Q^2}\cosh \lambda+R(\Omega_L+{2\Phi J_L\over Q^2})\sinh\lambda
\ .
\eea

\section{5D charged black strings}
In this section we find 5D black string solutions of the same theory (\ref{act}) which have the same near-horizon geometries as the 5D spinning black holes of the preceding section. The solutions are asymptotically flat in three, rather than four spatial directions and
translationally invariant along the fourth.
\subsection{Full solution and charges }
A 5D black string is  characterized by two magnetic charges obtained by integrating the two $U(1)$ field strengths over the $S^2$ surrounding the string.
We will fix the KK magnetic charge to unity.  This corresponds to having one Taub-NUT magnetic string.\footnote{Solutions with KK magnetic charge $p$ are related to $Z_p$ quotients of 5D black holes.} The second RR magnetic charge, which we denote $\Q$ below, counts both the number of
$K3$-wrapped D5-branes and parallel D1 branes which are equal due to the self-duality constraint we have imposed. The magnetic string worldvolume contains two conserved  $U(1)$ currents arising from the two 5D
bulk $U(1)$ gauge symmetries.   The general  black string solution has nonzero charge densities $q_{KK}$ and $q_{RR}$ along the string detected by long range electric fields at infinity. Hence we expect that the lowest-energy translationally and rotationally invariant configuration  is described by the three internal parameters $\Q,q_{KK}$ and $q_{RR}$. We also consider a momentum density along the string, but this is not an independent  fourth parameter as it can always be eliminated by a boost.

Defining \bea
\Omega_2&=&\sin\theta d\theta\wedge d\phi,\quad A_e=-{d{\hat t}\over {\hat r}+P}
\nn
\\
H&=&1+{P\over \hat{r}}\nn
\eea
the action (\ref{act})  has the two-parameter family of solutions
\bea \label{nhekstring2}
{ ds^2}&=&-H^{-2}d\hat{t}^2+H^2\left(d\hat{r}^2+\hat{r}^2(d\theta^2+\sin^2\theta d\phi^2) \right)\\
\nn&&+P^2\gamma[(d\hat{\psi}+\cos\theta d\phi+\alpha A_e)^2+(d\hat{y}+
\sqrt{1-\alpha^2}A_e)^2]\nn\\
F^{RR}_{(3)}&=&P^2 \tanh 2 \delta\left(dA_e \wedge (d\hat{y}+A_e)+\Omega_2\wedge (d\hat{\psi}+\cos\theta d\phi)\right)\nn\\&&
-{P^2\tanh 2 \delta \over  \cosh 2\delta }\left( -dA_e \wedge(d\hat{\psi}+\cos\theta d\phi)+\Omega_2\wedge(d\hat{y}-A_e) \right).\nn\eea
where $\alpha,~\gamma$ are related to $\delta$ by (\ref{alphagamma}).
Regularity at the poles requires
\be\hat{\psi}\sim \hat{\psi}+4
\pi .\ee
It can be verified that the RR 3-form is self-dual.
At large $\hat r$, $H\rightarrow1$ and the metric is
\be \label{nhring2}
{ds^2}=- d\hat{t}^2+d\hat{r}^2+\hat{r}^2(d\theta^2+\sin^2\theta d\phi^2) +P^2\gamma d\hat{y}^2
+P^2\gamma(d\hat{\psi}+\cos\theta d\phi)^2.
\ee
This is locally a product of an $S^1$ parameterized by $\hat\psi$ with 5-dimensional Minkowski space and can be viewed as a Kaluza-Klein (KK) compactification to five dimensions.\footnote{We expect that there is a more general solution in which the asymptotic radius of the KK $S^1$ can be varied arbitrarily, and that this is a special case in which the scalar sits at the attractor value throughout the geometry. As we are ultimately interested in the near-horizon behavior we will not consider this more general solution.}

Note that  the $S^1$ parameterized by $\hat \psi$  is fibered over the $S^2$ parameterized by $(\theta,\phi)$ in a manner indicating there is a string extending in the $\hat y$ direction carrying one unit of KK magnetic charge.
This string also carries D1 and D5 charges $Q_1$ and $Q_5$ given by
\be  Q_1=Q_5={1 \over 4\pi^2}\int_{S_3}F^{RR}_{(3)} =4P^2 \tanh 2 \delta \equiv \Q.\ee

In addition the string carries electric charge densities. To see this we dimensionally reduce to five dimensions along the $\hat \psi $ circle, with the ansatz \be ds^2_6=\eta^{-{1\over3}}ds^2_5+\eta(d{\hat \psi}+{\cal A})^2 \ee
which gives the 5D effective action
\bea
 {\cal S}_{5B} &=&{1\over2\pi^2}\int d^5x\sqrt{-g} \{R
-\ft13({\p \eta\over\eta})^2
-\ft1{6} \eta^{2\over3}\, (F_\3^\rr)^2
-\ft14 \eta^{4\over3}\, (\cF)^2
\}
\ \label{2b5d}
\eea
where $\cF=d{\cal A}$ and the KK scalar for the black string (\ref{nhekstring}) is given by
\bea \eta=P^2\gamma\ .\eea
The 5D two-form arising from $S^1$ reduction of the three-form 6D RR field is proportional to the 5D dual of $F_\3^\rr$ due to the self duality condition, and so does not explicitly appear in (\ref{2b5d}). More details of the 6D to 5D reduction can be found in \cite{Cremmer:1997ct} and \cite{Duff:1998cr}.
The KK gauge field can be read off of the $g_{\hat \psi \phi}$ and $g_{\hat \psi \hat t}$ terms in (\ref{nhekstring2}) and is
\be \cal A= \cos \theta d\phi+ \alpha A_e.\ee
The first term gives the magnetic string charge
\be q_{TN}=-{1\over4\pi}\int_{S^2}{\cal F} =1\ee
while the second gives a radial  electric field at large $\hat r$
\be {\cal F}_e=-{\alpha \over  \hat r^2}d\hat t \wedge d\hat r  .\ee
The proper density of KK electric charge on the string is hence
\be q_{KK}=-{1\over \pi^2} \int_{S^2\times R_y} \eta^{4\over3}*{\cal F}=-{\alpha\over\pi }(P^2\gamma)^{4\over3}.\ee
where $R_y$ is any interval along the $y$ direction with unit proper length, and $*$ is the five dimensional Hodge dual.
There is also an RR electric field at infinity.  The corresponding charge density is
 \be q_{RR}={1\over 4\pi^2} \int_{S^2\times R_y}  F^{RR}_{(3)}={\Q{\rm sech }2\delta\over 4\pi(P^2\gamma)^{{2\over3}}}.\ee
This electric field  is sourced by D1 branes wrapping the $S^1$ parameterized by ${\hat \psi}$ and smeared along the $\hat y$ direction of the string, as well as smeared  D5-branes wrapping $S^1\times K3$. The proper densities of these objects along the string are equal and both given by $q_{RR}$.
This string also has finite proper momentum and energy densities $ {\cal P}$ and
${\cal E}$. These are determined from the $1 \over \hat r$ corrections to the metric via the standard ADM formulae \cite{Lu:1993vt, Myers:1999psa, Harmark:2004ch}.
Using asymptotically Cartesian coordinates, the stress-energy tensor for a $p$-brane in $d$ dimensional spacetime is
\bea
T_{a b}&=&{1\over16\pi G_d}\oint d\Omega_{d-p-2} r^{d-p-2} n^i[\eta_{a b}(\p_i h^c_c+\p_ih^j_j-\p_jh^j_i)-\p_ih_{ab}]\label{stress}
\eea
where $n^i$ is a radial unit vector in the transverse space, and $h_{\mu\nu}=g_{\mu\nu}-\eta_{\mu\nu}\,.$ The labels $a, b=0,1\cdots p$ run over the world volume directions, while
$i,j$ denotes the transverse directions.
For the black string solution (\ref{nhekstring}), one first need to write the five dimensional Einstein frame metric in the canonical form, which amounts to the rescaling \be \hat{t}= \eta^{-{1\over6}}\tilde{t},\quad\hat{r}=\eta^{-{1\over6}}\tilde{r},\quad\hat{y}= \eta^{-{2\over3}}\tilde{y} \ee
Rewriting the metric (\ref{nhekstring}) in terms of tilded coordinates, and working out the stress tensor (\ref{stress}), one find
the energy and momentum densities
\be \label{ep} {\cal E}=T_{\tilde{t}\tilde{t}}={8\over\pi}P(P^2\gamma)^{1\over6},~~~ {\cal P}=T_{\tilde{t}\tilde{y}}=-{2\over\pi}(P^2\gamma)^{2/3} \sqrt{1-\alpha^2}\ .\ee
One could transform everything to a boosted frame with ${\cal P}=0$ but the resulting formulae are long and unilluminating.

In summary the supergravity solution (\ref{nhekstring2}) corresponds to the energy-momentum-density (\ref{ep}) brane configuration of the following table :
\begin{center}
\begin{tabular}{c|clcccccc}
 &Wrapping & 6 & 7& 8 & 9 & $u$ & $ \mathbb{R}_{y} $  \\ \hline
D5 &$\Q$& X & X& X & X &  & X  \\
D1 &$\Q$& & & & & & X  \\
TN &$1$& & & & &  &X \\
D5 &$q_{RR}$ & X & X& X & X & X &   \\
D1 &$q_{RR}$& & & & & X&  \\
KK &$q_{KK}$ & & & & &  X&\\
\end{tabular}
\end{center}
\medskip
where $\mathbb{R}_{y} $ denotes the common string direction, 6-7-8-9 are the $K3$ directions, the last three rows are proper wrapping densities and the last row denotes KK momentum around the $S^1$.

\subsection{Near-horizon limit}
The near horizon limit is taken by defining
\bea
\hat{r}&=&\epsilon r,\quad \hat{t}={P^2 t\over \epsilon},\\
\hat{\psi}&=&\psi+\alpha (y+ {P t\over \epsilon})\\
\hat{y}&=& \sqrt{1-\alpha^2}(y+ {P t\over \epsilon})
\eea
Taking the limit $\epsilon \to 0$
gives the near-horizon metric
\bea\label{nic}
{1\over P^2}ds^2 &=& -w_+ w_- + \gamma w_3^2+ \sigma_1^2 + \sigma_2^2  + \gamma \sigma_3^2 +2 \alpha \gamma w_3 \sigma_3
\eea
with the identification \be{\psi}\sim {\psi} +4\pi \ee
and three-form
\be \label{hfs}
 F^{RR}_{(3)} = { \Q \over4} \left( \sigma_1 \wedge \sigma_2 \wedge \sigma_3 + \frac{1}{2} \, w_+
\wedge w_-\wedge w_3  + {\rm sech} 2 \delta \, ( \sigma_1 \wedge \sigma_2 \wedge w_3 +
\frac{1}{2} \, w_+ \wedge w_-\wedge
\sigma_3 )\right)
\ee

This general near-horizon geometry is fully described by two parameters which can be taken to be $\delta$ and $P$.  The asymptotically flat solution on the other hand has a three-parameter generalization in which $q_{KK}$ and $q_{RR}$ are separately varied. These all reduce, however, in the near-horizon limit to the same two parameter family given above. This corresponds, in the dual picture, to radial RG flows which differ in the UV but reach the same IR fixed point. This observation agrees with the analysis of \cite{Compere:2010fm}, where  general black string solutions in the minimal 5D supergravity with a single $U(1)$ were studied. In that case, the radial electric field is always scaled away in the near-horizon region, \footnote{For rotating black strings, a different decoupling limit was discussed in \cite{Compere:2010uk}, under which there is also a warped horizon, but the $SU(2)$ symmetry is broken.} which is therefore  always AdS$_3$.\footnote{The fact that one of the two electric fields scales away in the near-horizon region is related to the fact that the near horizon $SL(2,R)_L\times U(1)_R$ isometry contains a left but not a right scaling symmetry.}  We see here that in order to get a rotationally-invariant near-horizon warped AdS$_3$ two $U(1)$s are required.
\section{NHEK = black string near-horizon}
The near-horizon black string geometry (\ref{nic}), (\ref{hfs}) is locally the same as the NHEK geometry (\ref{nicemetric}), (\ref{hf}) with the identification \be M=12P^2. \ee
Globally the NHEK geometry also has the identification (\ref{id}) of the $y$ coordinate. In  terms of the hatted coordinates of the full black string solution (\ref{nhekstring2}) this identification is
    \be \label{idd} \hat y \sim \hat y+{4 \pi^2T_R\sqrt{1-\alpha^2} m },~~~~~\hat \psi \sim \hat \psi+{2 \pi}(2\alpha\pi T_R+\Theta)m+4\pi n. \ee
In other words, the 5D black string is wrapped around a compactified  $\hat y$ circle and has boundary conditions twisted by a rotation along $\hat \psi$.

In conclusion, the dual of the 5D extreme spinning black holes is identified as the low energy limit of the D-brane configuration in table 1 wrapped on a circle with $\hat \psi$-twisted boundary conditions.

\section*{Acknowledgements}
We are grateful to Geoffrey Compere, Alessandra Gnecchi,  Monica Guica, Shamit Kachru, Josh Lapan, Juan Maldacena and Alex Maloney for useful conservations.  This work was supported in part by
DOE grant DE-FG02-91ER40654 and the Harvard Society of Fellows.

\end{document}